# Structural Breaks in the Mexico's Integration into the World Stock Market

**Mohamed El Hédi AROURI**

LEO, Université d'Orléans and EDHEC, France

Mohamed.arouri@univ-orleans.fr

**Jamel JOUINI**

FSEGN and LEGI-EPT

Université 7 Novembre de Carthage, Tunisia

GREQAM, Université de la Méditerranée, France

Jamel.Jouini@fsegn.rnu.tn

**Abstract**

This article investigates the evolution of the Mexican stock market integration into the world market. First, we estimate the time-varying Mexican degree of market integration using an international conditional version of the CAPM with segmentation effects. Second, we study the structural breaks in this series. Finally, we relate the obtained results to important facts and economic events.

# 1- Introduction

Studying the integration of a domestic market into the world market is an empirical question that has decisive impact on a number of issues addressed by finance literature. If capital markets are integrated, investors face common and country-specific risks, but price only common risk because country-specific risk is diversified internationally. In this case, the same asset pricing relationships apply in all countries. In contrast, when markets are segmented the asset pricing relationship varies across countries and returns would be determined by domestic factors. When markets are partially segmented, investors face both common and country-specific risks and price them both. In this case, returns should be determined by a combination of local and global factors.

Empirical papers investigating stock market integration have been mainly limited to developed markets [De Santis and Gerard (1997), De Santis *et al.* (2003), and Hardouvelis *et al.* (2006)]. These studies support the integration hypothesis of developed markets. Recently, some papers have tented to focus on emerging markets, in particular Asian and Latin American markets [Gérard *et al.* (2003), Bekaert *et al.* (2005), and Carrieri *et al.* (2007)]. The results of these studies are heterogeneous, but conclude that emerging markets are partially segmented and their degrees of integration are time-varying.

In this article, we develop and test an international conditional capital asset pricing model (CAPM) with segmentation effects in order to infer a time-varying measure of market integration. Unlike most previous works which study market integration in cross sections of countries, we follow Adler and Qi (2003) and investigate the issue thought a longitudinal study of a single market, Mexico, over the last twenty years. Mexico is the biggest Latin American market almost fully accessible to foreign investors. In fact, in the last two decades foreign investment barriers were reduced, country funds were introduced and depository receipts (DR) were listed in order to improve the integration of Mexico into the world market. Integration should drive to a lower cost of capital, bigger investment opportunities and higher economic growth [Bekaert and Harvey (2003)]. Studying the Mexican stock market leads to a better view of the integration process. Furthermore, we test for structural breaks in the obtained degree of integration and try to explain changes by important facts and economic events.

Section 2 presents the methodology. Section 3 describes the data and reports the main empirical results. Concluding remarks are in section 4.

## 2- Methodology

The CAPM predicts that the expected excess return on an asset is proportional to its systematic risk. Under integration, an international conditional version of the CAPM can be written as:

$$E(R_{it}^l / \Omega_{t-1}) = \delta_{w,t-1} Cov(R_{it}^l, R_{wt} / \Omega_{t-1}), \quad \forall i, \quad (1)$$

where $R_{it}^l$ and $R_{wt}$ are respectively the excess returns on asset $l$ in country $i$ and on the world market, $\delta_{w,t-1}$ is the price of world market risk. Expectations are taken with respect to the set of information variables $\Omega_{t-1}$.

Conversely, under segmentation, the domestic CAPM holds:

$$E(R_{it}^l / \Omega_{t-1}) = \delta_{i,t-1} Cov(R_{it}^l, R_{it} / \Omega_{t-1}), \quad \forall l, i, \quad (2)$$

where $R_{it}$ is the excess return on market portfolio of country $i$ and $\delta_{i,t-1}$ is the price of domestic risk.

At the national level, (2) becomes:

$$E(R_{it} / \Omega_{t-1}) = \delta_{i,t-1} Var(R_{it} / \Omega_{t-1}), \quad \forall i. \quad (3)$$

However, recent studies suggest that returns should be influenced by both global and local factors [Bekaert and Harvey (1995), and Carrieri *et al.* (2007)]. In this partially segmented framework, the returns are given by:

$$E(R_{it} / \Omega_{t-1}) = \varphi_{i,t-1} \delta_{w,t-1} Cov(R_{it}, R_{wt} / \Omega_{t-1}) + (1 - \varphi_{i,t-1}) \delta_{i,t-1} Var(R_{it} / \Omega_{t-1}), \quad \forall i, \quad (4)$$

where $\varphi_{i,t-1}$ is a measure of market integration.

If $\varphi_{i,t-1}=0$, only domestic variance is priced and the market $i$ is segmented. Whereas, if $\varphi_{i,t-1}=1$, only the world risk is priced and the market $i$ is integrated. Finally, if $0<\varphi_{i,t-1}<1$, the market $i$ is partially segmented.

Next, consider the econometric methodology. Equation (4) has to hold for both Mexican and world markets. Under rational expectations, we can write:

$$\begin{aligned} R_{m,t} &= \varphi_{t-1}\delta_{w,t-1}h_{m,w,t} + (1-\varphi_{t-1})\delta_{i,t-1}h_{m,t} + \varepsilon_{m,t}, \\ R_{w,t} &= \delta_{w,t-1}h_{wt} + \varepsilon_{w,t}, \end{aligned} \qquad (5)$$

where $\varepsilon_t = (\varepsilon_{m,t},\varepsilon_{w,t})'/\Omega_{t-1} \sim N(0,H_t)$, $H_t$ is the $(2\times 2)$ conditional covariance matrix of returns, $h_{m,w,t}$ is the conditional covariance between Mexican and world markets, $h_{m,t}$ and $h_{w,t}$ are respectively the conditional variance of Mexican and world markets.

$H_t$ is given by:

$$H_t = C'C + aa'*\varepsilon_{t-1}\varepsilon'_{t-1} + bb'*H_{t-1}, \qquad (6)$$

where $C$ is a $(2\times 2)$ lower triangular matrix and $a$ and $b$ are $(2\times 1)$ vectors.

Finally, we follow previous works to specify the evolution of prices of risk. These prices are modelled as a positive function of information variables: $\delta_{w,t-1} = \exp(\kappa'_w Z_{t-1})$ and $\delta_{i,t-1} = \exp\left(\kappa'_i Z^i_{t-1}\right)$, where $Z$ and $Z^i$ are respectively a set of global and local variables included in $\Omega_{t-1}$. As in Hardouvelis *et al.* (2006), the time-varying function $\varphi_{i,t-1}$ is conditioned on a set of variables that affect market integration: $\varphi_{i,t-1} = 1 - Exp(-(\delta'_i Z^*_{i,t-1})^2)$, where $Z^*_{i,t-1}$ is a set of variables expected to be correlated with market integration. By construction $0 \le \varphi_{i,t-1} \le 1$, $\varphi(\pm\infty)=1$ and $\varphi(0)=0$. We take into account these features in the construction of variables. Precisely, we will assume that deviations of variables from zero, independent of their sign, reduce the degree of integration. The quasi-maximum likelihood (QML) method is used to estimate the model.

Once the time-varying degree of market integration becomes available, we test for structural breaks. Let $y_t$ be the degree of integration. We consider the following mean-shift model with $m$ breaks, $(T_1, T_2, ..., T_m)$:[1]

$$y_t = \mu_j + u_t, \qquad t = T_{j-1} + 1, ..., T_j, \qquad (7)$$

for $j = 1, ..., m+1$, $T_0 = 0$ and $T_{m+1} = T$. $\mu_j$ are the regression coefficients with $\mu_i \neq \mu_{i+1}$ $(1 \leq i \leq m)$, and $u_t$ is the error term. The estimation method developed by Bai and Perron (1998) is based on the ordinary least-squares principle. It consists in estimating the regression coefficients $\mu_j$,[2] and the break dates $(T_1, T_2, ..., T_m)$ under the condition that $T_i - T_{i-1} \geq [\varepsilon T]$, where $\varepsilon$ is an arbitrary small positive number and [.] denotes integer part of argument.

Bai and Perron (2003) propose a test-based selection procedure to estimate the number of breaks. Indeed, they suggest to first look at the results of tests $UD \max F_T$ or $WD \max F_T$,[3] to see if at least one structural break exists. The number of breaks is then determined based upon a sequential examination of a test $\sup F_T (l+1/l)$.[4] We then choose $m$ break dates such that the test $\sup F_T (l+1/l)$ is not significant for any $l \geq m$.[5]

**3- Data and Results**

*Data*

We use monthly stock returns for Mexico and world markets over the period January 1988–February 2008. Returns include dividend yields and are computed in excess of the 30-day Eurodollar deposit rate. In order to preserve comparability with previous studies, the choice of global, local and integration information variables is mainly drawn from previous works. The set of global information includes a constant, the MSCI world dividend price ratio in excess of the 30-day Eurodollar deposit rate (WDY), the change in the US term premium spread (DUSTP), the US default premium (USDP) and the change on the one month Eurodollar deposit rate (DWIR). The set of local information includes a constant, the Mexican dividend price ratio in excess of the local short-term interest rate (LDY), the change in the Mexican short-term interest rate (DLIR) and the change in industrial production (DIP). The set of integration variables includes a constant, the difference between the world and the Mexican dividend yields (DDY), the difference between the G7 and the Mexican real short-term interest rates (DIR) and the volatility of the exchange rate *vis-à-vis* the US dollar (VER). Descriptive statistics for returns and information variables, available upon request, support the GARCH parameterisation we use, and show in particular that our proxy of the information set contains no redundant variables.

---

[1] A look at Figure 1 suggests that the series may be affected by structural breaks with potential mean-shifts.
[2] The estimated coefficient $\hat{\mu}_j$ measures the average integration degree in the regime $j$.
[3] The hypothesis of no break versus an unknown number of changes given a maximum number of breaks $M$ for $m$ is tested.
[4] It tests the null hypothesis of $l$ breaks against the alternative that an additional break exists.
[5] For the application of the test procedure, see Bai and Perron (2003), and Jouini and Boutahar (2005).

*Time-varying degree of integration*

Table 1 contains parameter estimates and diagnostic tests. The ARCH and GARCH coefficients reported in Panel B are significant. Panel A shows the mean equation parameter estimates, Panel C presents standardized residual diagnostics and Panel D reports some specification tests. Most information variables are significant. The world and domestic prices of risk are significantly time-varying. On average, they are respectively equal to 3.47 and 2.34. Thus, Mexico is partially integrated into the world market: both global and local risks are priced. Diagnostics of standardized residuals show that compared to returns series, the non-normality is reduced and there is no residual autocorrelation.

Wald test shows that the Mexican degree of integration into the world market is significantly time-varying (Figure 1). The average degree of integration is 57%. Mexico was segmented at the beginning of our sample with a degree of integration on average less than 50%. This market has recently became highly integrated and its degree of integration has exceeded 75% in the last two years.[6] This result is intuitive given the removal of all restrictions on foreign direct purchases of non bank stocks and DR listings since mid-1990s and, in particular, the degree of US investor participation in Mexican stocks. Next, we study structural breaks in this degree of integration.

*Structural Breaks*

Table 2 summarizes the results of the structural break procedure for $M = 5$ and $\varepsilon = 0.10$. Four break dates are obtained: December 1992, December 1994, May 2001 and December 2005.[7] The detected breaks can be related to important economic events. The North-American Free Trade Agreement (NAFTA) negotiation and reduction of capital movement barriers from 1990 to 1994 improved the Mexican market integration. However, the peso was fixed to US dollar, which was incompatible with the high inflation and affected the Mexican economy competitiveness. As a result, Mexico's integration decreased in 1995 due to the crisis and peso devaluation.

---
[6] Adler and Qi (2003), and Carrieri *et al.* (2007) have shown a higher integration of Mexico in the recent period.

The degree of integration has increased since 2001. Several factors may justify a high integration of the Mexican market into the world in the recent period: the improvement of economic and social stability, the institutional reforms, the liberal policies that implied a commercial and financial "dereglementation" of the economic activity and privatization. The Mexican stock market has been particularly distinguished by a strong growth of its capitalization. In fact, the later increased by 23% in 2005, and 44% of securities were detained by foreign investors. Moreover, its ties with the USA explain the increase of the foreign direct investments that reached more than 11 billions US $ in 2000 against 4,4 billions US $ in 1993. Finally, several multinational enterprises have chosen Mexico to extend their activities in the United States.

## 4- Conclusion

In this paper, the question of estimating the Mexican degree of integration has been subjected to a meticulous examination using an international conditional version of the CAPM with segmentation effects. The results show that the Mexico is partially integrated into world market and its degree of integration is time-varying. The application of a structural break procedure allows identifying four break dates which can be related to important economic facts.

---

[7] These dates are illustrated in Figure 1. They are precisely estimated since the corresponding confidence intervals cover a few months before and after.

## Table 1: QML estimates of the model

**Panel A**: Mean equations

*(a) Price of world risk*

|  | Const. | WDY | DUSTP | USDP | DWIR |
|---|---|---|---|---|---|
| **Price of market risk** | 0.354* | 1.012* | -0.655 | 0.679** | -0.867*** |
|  | (0.023) | (0.234) | (1.098) | (0.367) | (0.411) |

*(b) Price of Mexican risk*

|  | Const. | LDY | DLIR | DIP |
|---|---|---|---|---|
| **Price of Mexican risk** | 0.405** | -1.156** | -0.044*** | -0.542 |
|  | (0.189) | (0.427) | (0.031) | (1.067) |

*(c) Degree of Mexican market integration*

|  | Const. | DDY | DIR | VER |
|---|---|---|---|---|
| **Degree of integration** | 0.201* | 0.312** | 1.944*** | -0.493** |
|  | (0.023) | (0.142) | (1.114) | (0.226) |

**Panel B**: GARCH process

|  | Mexico | World |
|---|---|---|
| **a** | 0.103* | 0.133* |
|  | (0.045) | (0.035) |
| **b** | 0.597* | 0.821* |
|  | (0.201) | (0.114) |

**Panel C**: Residual diagnostics

|  | U.S. | World |
|---|---|---|
| *Skewness* | -0.477* | -0.417* |
| *Kurtosis* | 1.549* | 1.164* |
| *J.B.* | 33.393* | 34.122* |
| *Q(12)* | 7.352 | 12.628 |

**Panel D**: Specification tests

| Null hypothesis | $\chi^2$ | df | p-value |
|---|---|---|---|
| *Is the price of world risk constant?* <br> $H_0: \delta_{w,j}=0 \quad \forall j>1$ | 47.56 | 4 | 0.000 |
| *Is the price of Mexican risk constant?* <br> $H_0: \delta_{d,j}=0 \quad \forall j>1$ | 7.75 | 3 | 0.043 |
| *Is the degree of integration constant?* <br> $H_0: \varphi_j=0 \quad \forall j>1$ | 76.10 | 3 | 0.000 |

*Note: *, ** and *** Denote significance at 1%, 5% and 10%.. QML robust standard errors are in parentheses. In order to preserve space, estimates of C is not reported.*

## Table 2: Structural break identification

| Break Dates | $\hat{T}_1$ | $\hat{T}_2$ | $\hat{T}_3$ | $\hat{T}_4$ |
|---|---|---|---|---|
|  | 1992:12 | 1994:12 | 2001:5 | 2005:12 |
|  | (1992:7-1993:2) | (1994:7-1995:6) | (2000:12-2001:7) | (2005:7-2006:4) |
| Regression Coefficients | $\hat{\mu}_1$ | $\hat{\mu}_2$ | $\hat{\mu}_3$ | $\hat{\mu}_4$ |
|  | 0.471 | 0.601 | 0.504 | 0.651 |
|  | (0.006) | (0.014) | (0.006) | (0.010) |
|  | $\hat{\mu}_5$ |  |  |  |
|  | 0.790 |  |  |  |
|  | (0.015) |  |  |  |

*Note: The 95% confidence intervals for the break dates and the standard errors (robust to serial correlation) for coefficients are in parentheses.*

**Figure 1**: **Time-varying Mexico's degree of integration**

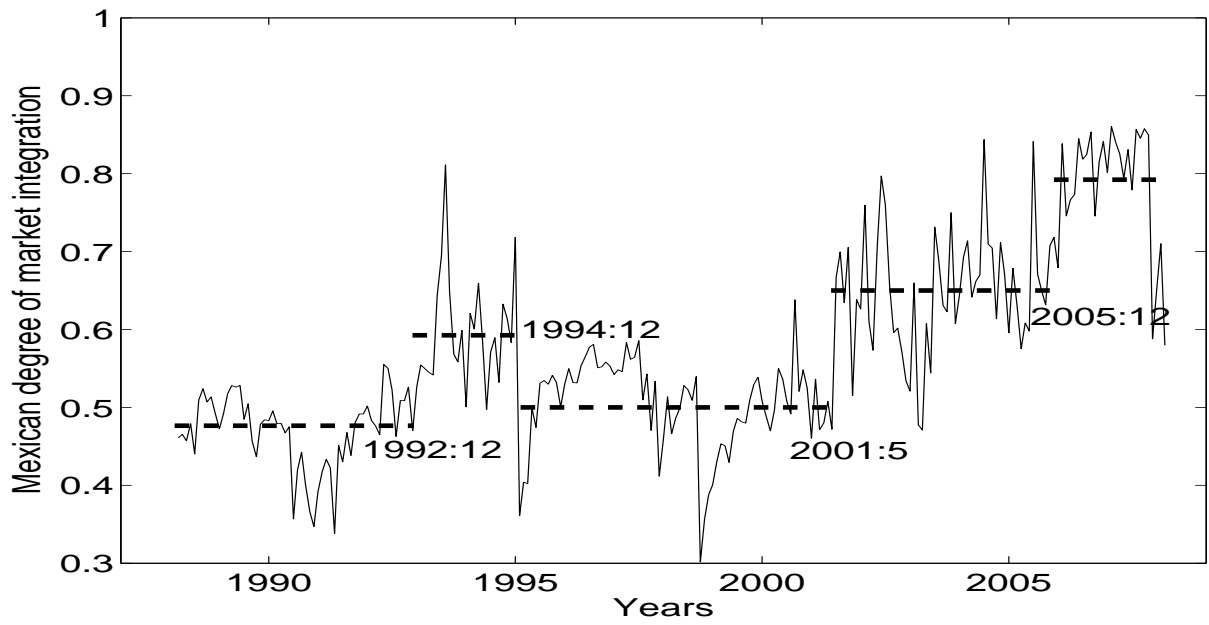